\documentclass[conference]{IEEEtran}
\IEEEoverridecommandlockouts
\usepackage{cite}
\usepackage{amsmath,amssymb,amsfonts}
\usepackage{algorithmic}
\usepackage{graphicx}
\usepackage{textcomp}
\usepackage{xcolor}
\usepackage[letterpaper, left=0.625in, right=0.625in, bottom=1in, top=0.75in]{geometry}
\usepackage{caption}
\usepackage{subcaption}
\def\BibTeX{{\rm B\kern-.05em{\sc i\kern-.025em b}\kern-.08em
    T\kern-.1667em\lower.7ex\hbox{E}\kern-.125emX}}
    
\usepackage{hyperref}
\hypersetup{
    colorlinks=true,
    linkcolor=blue,
    filecolor=magenta,      
    urlcolor=cyan,
    pdftitle={Overleaf Example},
    pdfpagemode=FullScreen,
    }

\usepackage[T1]{fontenc}
\usepackage{caption}
\usepackage[font=footnotesize]{subcaption}
\usepackage{graphicx}
\usepackage{siunitx}

\begin{document}
\title{GeoTyper: Automated Pipeline from Raw scRNA-Seq Data to Cell Type Identification\\

}

\author{\IEEEauthorblockN{Cecily Wolfe}
\IEEEauthorblockA{\textit{School of Data Science} \\
\textit{University of Virginia}\\
Charlottesville, VA \\
cew4pf@virginia.edu}
\and
\IEEEauthorblockN {Yayi Feng}
\IEEEauthorblockA{\textit{School of Data Science} \\
\textit{University of Virginia}\\
Charlottesville, VA \\
yf7qq@virginia.edu}
\and
\IEEEauthorblockN{David Chen}
\IEEEauthorblockA{\textit{School of Data Science} \\
\textit{University of Virginia}\\
Charlottesville, VA \\
dzc5ta@virginia.edu}
\and
\IEEEauthorblockN{Edwin Purcell}
\IEEEauthorblockA{\textit{School of Data Science} \\
\textit{University of Virginia}\\
Charlottesville, VA \\
jzd6af@virginia.edu}
\and
\IEEEauthorblockN{Anne Talkington}
\IEEEauthorblockA{\textit{Department of Biomedical Engineering} \\
\textit{University of Virginia}\\
Charlottesville, VA \\
vmp8nt@virginia.edu}
\and
\IEEEauthorblockN{Sepideh Dolatshahi}
\IEEEauthorblockA{\textit{Department of Biomedical Engineering} \\
\textit{University of Virginia}\\
Charlottesville, VA \\
sd8us@virginia.edu}
\and
\IEEEauthorblockN{Heman Shakeri}
\IEEEauthorblockA{\textit{School of Data Science} \\
\textit{University of Virginia}\\
Charlottesville, VA \\
hs9hd@virginia.edu}
}

\maketitle

\begin{abstract}
The cellular composition of the tumor microenvironment can directly impact cancer progression and the efficacy of therapeutics. Understanding immune cell activity, the body's natural defense mechanism, in the vicinity of cancerous cells is essential for developing beneficial treatments. Single cell RNA sequencing (scRNA-seq) enables the examination of gene expression on an individual cell basis, providing crucial information regarding both the disturbances in cell functioning caused by cancer and cell-cell communication in the tumor microenvironment. This novel technique generates large amounts of data, which require proper processing. Various tools exist to facilitate this processing but need to be organized to standardize the workflow from data wrangling to visualization, cell type identification, and analysis of changes in cellular activity, both from the standpoint of malignant cells and immune stromal cells that eliminate them. We aimed to develop a standardized pipeline (GeoTyper, https://github.com/celineyayifeng/GeoTyper) that integrates multiple scRNA-seq tools for processing raw sequence data extracted from NCBI GEO, visualization of results, statistical analysis, and cell type identification. This pipeline leverages existing tools, such as Cellranger from 10X Genomics, Alevin, and Seurat, to cluster cells and identify cell types based on gene expression profiles. We successfully tested and validated the pipeline on several publicly available scRNA-seq datasets, resulting in clusters corresponding to distinct cell types. By determining the cell types and their respective frequencies in the tumor microenvironment across multiple cancers, this workflow will help quantify changes in gene expression related to cell-cell communication and identify possible therapeutic targets.
\end{abstract}

\begin{IEEEkeywords}
scRNA-Seq, Cancer, Pipeline, NCBI GEO, Seurat, Cell Type Identification
\end{IEEEkeywords}

\section{Introduction}
Immune system evasion is one of the accepted hallmarks of cancer \cite{b1}, but the role of the immune system offers opportunities for leveraging the body’s natural defense mechanisms. In order to promote the development of cancer treatments, an understanding of immune cell activity, the body’s natural defense mechanism, in the vicinity of cancerous cells is essential. The tumor microenvironment consists of various cell types, including malignant, stromal, and immune cells, and these different cell types communicate through ligand-receptor interaction \cite{b2}. Ligand-receptor interaction constitutes one of the foundations of developing therapeutics for targeting cell-cell communications \cite{b2}. The investigation of cell-cell communication offers insights into mechanisms such as tumorigenesis, tumor proliferation, therapy resistance, and other treatment complications \cite{b1}. The impact of cell-cell communication on patient treatment and outcome can be leveraged to develop therapeutics that target the relationship between cell types \cite{b2}. Yet, the general response rate to these therapeutics is limited due to the complex nature of cell-cell communications \cite{b3}. Thus, it is essential to better understand and target the interactions between cells for promising therapeutic outcomes.

Single cell RNA sequencing (scRNA-seq) enables the examination of gene expression on an individual cell basis and details of heterogeneity of the cellular composition within the tumor and surrounding tissue, providing crucial information regarding cell-cell communication in the tumor microenvironment \cite{b2}. This novel technique provides insights into the heterogeneity of cellular composition by generating large amounts of data that require proper processing. There are many existing tools that can process scRNA-seq data such as 10X Genomics Cellranger Count, Alevin, and Seurat, and there are various cell identification techniques including SingleR, scCATCH, and an adapted version of the neural network ACTINN. Yet, there is a lack of a streamlined workflow that runs from extracting raw scRNA-seq data from public repositories such as the National Center for Biotechnology Information’s Gene Expression Omnibus (NCBI GEO) to identifying cell types of the data. The objective of this project was to develop a user-friendly and standardized pipeline that allows users who have either a background in Immunology or Computational Biology to be able to process data more quickly and efficiently. This pipeline has the potential to facilitate data processing in cancer research. 

\section{Literature Review}

Single cell RNA sequencing emerged just over a decade ago, with the first paper detailing the technique published by Tang et al. in 2009 \cite{b4}\cite{b5}. The number of scRNA-seq methods continues to expand, with at least six distinct approaches now commonly in use \cite{b6}. The differences in the raw data generated and the increased granularity in genomic information offered by scRNA-seq resulted in the development of numerous computational tools to help process and analyze the data, with an emphasis on automatic identification of cell types.

In light of many options, researchers have proposed multiple approaches to systematize analysis. Vieth et al. explored thousands of combinations of scRNA-seq library preparation, mapping, imputation, normalization, and differential expression methods, and concluded that tools used to normalize data had the biggest impact on pipeline performance with respect to quantifying changes in gene expression across cell types \cite{b7}. Luecken and Theis implemented best practices for pre-processing and downstream analysis by integrating tools based in both R and Python \cite{b8}. Slovin et al. proposed four step-by-step pipelines, from producing data in the laboratory to cell type identification and differential gene expression with platforms based in R and Python, including Seurat, Scanpy, Monocle, and gf-icf \cite{b9}.

Cell type identification in particular is essential for drawing biological conclusions from data. This is a non-trivial problem since it relies on extensive knowledge of the specific biological system, which is complicated by heterogeneity in expression of canonical markers. In a review published in 2021, Xie et al. \cite{b10} compared the performance of thirty-two automatic methods, including Seurat, an R package originally developed by Satija et al. \cite{b11}, and ACTINN (Automatic Cell Type Identification using Neural Networks) developed by Ma and Pelligrini \cite{b12}, scCATCH \cite{b13}, and SingleR \cite{b14}, the methods adapted for our own research. When evaluating approaches using either an eager learning (i.e., classifying test data based on pre-annotated training data), lazy learning (i.e., nearest neighbors), or marker learning (i.e., using canonical marker genes recognized based on empirical studies to serve as attributes unique to given cell types), Xie et al. concluded that the best packages in each category achieved similar results in terms of accuracy, F1-score, and speed \cite{b10}.

As all of these papers note, the field of scRNA-seq computational biology is still largely in its infancy, and documentation and comparative analysis have not kept pace with the explosion of available tools. The range of approaches makes an automated, standardized, and user-friendly workflow an attractive, yet elusive, analytical aid for studying subjects such as cell-cell communication and cancer immunotherapy. Despite the innovative approaches described above, there does not exist a widely accepted, end-to-end computational pipeline encompassing raw scRNA-seq data processing, quality control, data normalization, and cell type identification for users with varying levels of expertise in biological research and computational methods.

\section{Data Description}
We used several publicly available datasets to develop and hone our pipeline, focusing on those relevant to the immune system and cancer. Two datasets, containing peripheral blood mononuclear cells (PBMCs), or specialized immune cells \cite{b15}, and samples from several patients with lung cancer \cite{b16}, respectively, came from the 10X Genomics archive. Data from three patients with lymphoma were provided via NCBI GEO by Cerapio et al. \cite{b17} \cite{b18} as part of their study on differential T cell behavior across cancer types. All data were in the form of FASTQs, a text-based file format containing unique IDs for each cell, nucleotide sequences, and ASCII-encoded quality scores for each nucleotide (which represent the probability a given base is incorrectly identified). These FASTQ files were paired with each other, with one FASTQ file in a set containing unique nucleotide barcodes attached to each cell during scRNA-seq library preparation, and the other containing the nucleotide sequence from the biological sample.

\section{Methodology}

\subsection{Overview}\label{AA}

The pipeline consists of three major components: Data Processing, Cell Type Identification, and Visualization. The first step, Data Processing, involves using Cell Ranger or Alevin, depending on the type of machine used to collect the data. These data processing methods generate gene expression matrices, which contain counts of gene expressions on an individual cell basis. Next, Seurat is used to cluster the cells and visualize them using dimensionality reduction. Lastly, eager learning and marker learning methods such as PanglaoDB, singleR, scCATCH, and ACTINN, are used to assign cell types to  each cluster. 

\begin{figure}[htbp]
\centerline{\includegraphics[scale=0.4]{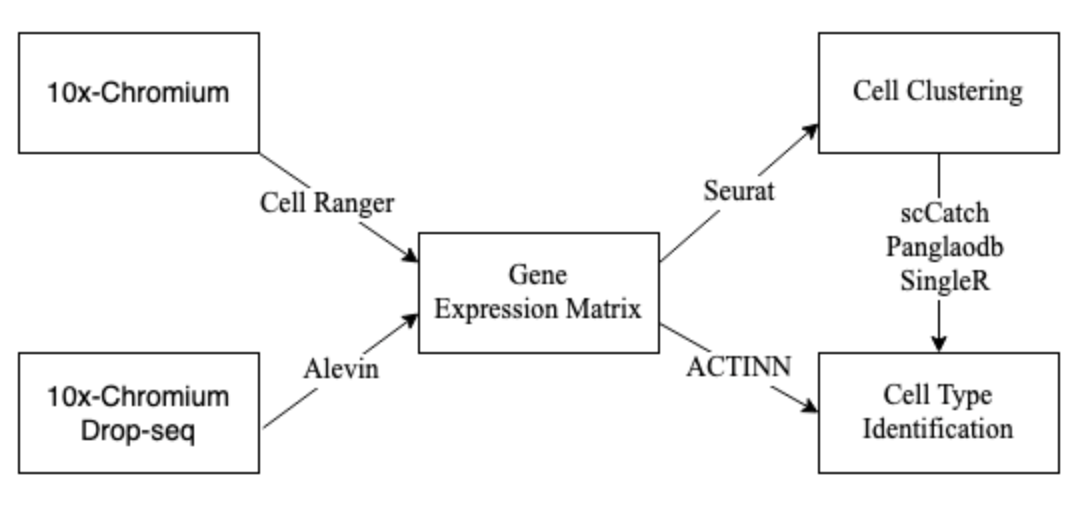}}
\caption{The pipeline can be visualized using boxes that represent files and arrows that represent programs.}
\label{fig}
\end{figure}

\subsection{Data Processing}
To retrieve the raw scRNA-seq data from NCBI GEO, we implemented the package “sratoolkit” \cite{b19}\cite{b20}, which allows users to specify the SRR run ID from SRA Run Selector in the slurm file to obtain the FASTQ files they would like to extract. The files would be downloaded and split into two zipped FASTQs – one for the UMIs or cell barcodes, and one with the actual cDNA.
The Cellranger Count pipeline from 10X Genomics aligns sequencing reads in FASTQ files to a reference transcriptome in order to generate a feature barcode matrix that serves as the input for the downstream analysis by Seurat.  For data generated using the 10X Chromium method, the most common approach is to use 10X Cellranger Count to carry out the alignment and feature counting. However, 10X Cell Ranger only supports 10X Chromium, and thus an alternative and more flexible approach is to use Alevin, which supports 10X Chromium and Drop-seq derived data. Similarly, Alevin generates a cell-by-gene count matrix given the reference transcriptome and the raw read files, which are constructed by processing the FASTQ files. Through both methods, an input of raw sequencing data can be transformed into a count matrix which has rows for each individual cell and columns for the degree of genes expressed within each cell. While Alevin is less computationally expensive to run compared to Cell Ranger, the performance of Alevin is highly variable in terms of exhibiting different behavior for genes per cell across different datasets \cite{b21}.

\subsection{Seurat Analysis}
Seurat is a standardized workflow for processing sc-RNA seq data including procedures such as selecting and filtering cells based on quality control (QC) metrics, normalizing and scaling data, generating clusters of cells, and running non-linear dimensional reduction. The QC metrics used are cells with unique feature counts over 4,500 or less than 200 are filtered and cells with the threshold of 25\% or two standard deviations above the mean, whichever is stricter. This filters out data points that contain more or less total counts of RNA reads than expected for a normal cell, indicating this “cell” may actually be multiple cells or cell fragments. The data are normalized using a log-transformed global-scaling normalization method “LogNormalize" after removing unwanted cells and scaled using a linear transformation prior to implementing dimensional reduction techniques such as Principal Component Analysis (PCA). Next, Seurat applies a graph-based cluster approach that constructs a K-Nearest Neighbors (KNN) graph based on Euclidean distance, refines the edge weights between any two cells based on Jaccard’s similarity, and groups cells together using modularity optimization techniques \cite{b22}. The clusters are visualized and explored using a non-linear dimension reduction algorithm, Uniform Manifold Approximation and Projection (UMAP), which permits the visualization of the clusters in a lower-dimensional space. UMAP, and the similar dimension reduction, t-distributed stochastic neighbor embedding (t-SNE), help visualize clusters of similar cells, as well as compare cell type identification methods based on labels assigned to different cells.

\subsection{Identifying Cell Types of the Clusters generated by Seurat using PanglaoDB}
Since the Seurat algorithm does not identify cell types for the clusters, we developed a novel approach to extract information from PanglaoDB, a database for scRNA sequencing experiments, and apply it to cell type identification \cite{b23}. We leveraged the top five canonical markers for each cluster as identified by Seurat and matched the markers to the markers found in PanglaoDB along with their respective cell types. The issue with this approach is that one canonical marker could be associated with multiple cell types; thus, we used the maximum difference between the sensitivity (how frequently this marker is expressed in cells of this particular cell type) and specificity (how frequently this marker is NOT expressed in cells of this particular cell type) as the metric to determine the most likely cell type for that particular canonical marker.

\subsection{Identifying Cell Types using scCATCH}
Single-cell Cluster-based automatic Annotation Toolkit for Cellular Heterogeneity, or scCATCH, is a marker learning method where cell types of clusters are annotated based on evidence-based score by matching the potential canonical markers with known cell markers in tissue-specific cell taxonomy reference database CellMatch \cite{b13}. As part of the package, scCATCH includes lists of canonical markers, or genes characteristically expressed at higher levels by certain types of cells than others, to discern which cells exhibit gene expression patterns associated with certain cell types \cite{b10}.

\subsection{Identifying Cell Types using ACTINN}
ACTINN (Automated Cell Type Identification using Neural Networks) is an eager learning method, which involves feeding a neural network with labeled data and having it predict the cell types of new and unseen data. ACTINN leverages a neural network of 3 hidden layers with 100, 50, and 25 nodes in the first, second, and third hidden layers, respectively \cite{b12}. In its current state, our implementation of ACTINN can classify cells as one of four types (B cell, monocyte, natural killer (NK) cell, T cell) or as unknown. Since the labeled human gene expression training data that the developers of ACTINN provided only contained several cell types, ACTINN only recognizes and assigns cells to those particular types \cite{b12}.

\subsection{Identifying Cell Types using SingleR.}
SingleR uses multiple rounds of analysis with Spearman correlation, or the extent to which variables are monotonically related, to match cells to the nearest known cell type from a reference dataset of pure cell types \cite{b14}. This iterative process leverages data from annotated bulk (as opposed to single cell) RNA sequencing in order to assign a cell type to each cell individually, rather than first clustering the cells and then assigning cell types to each cluster based on shared characteristics \cite{b14}.

\begin{figure*}	
\vspace*{+.127cm}
\centering
	\begin{subfigure}[t]{3in}
		\includegraphics[width=3in]{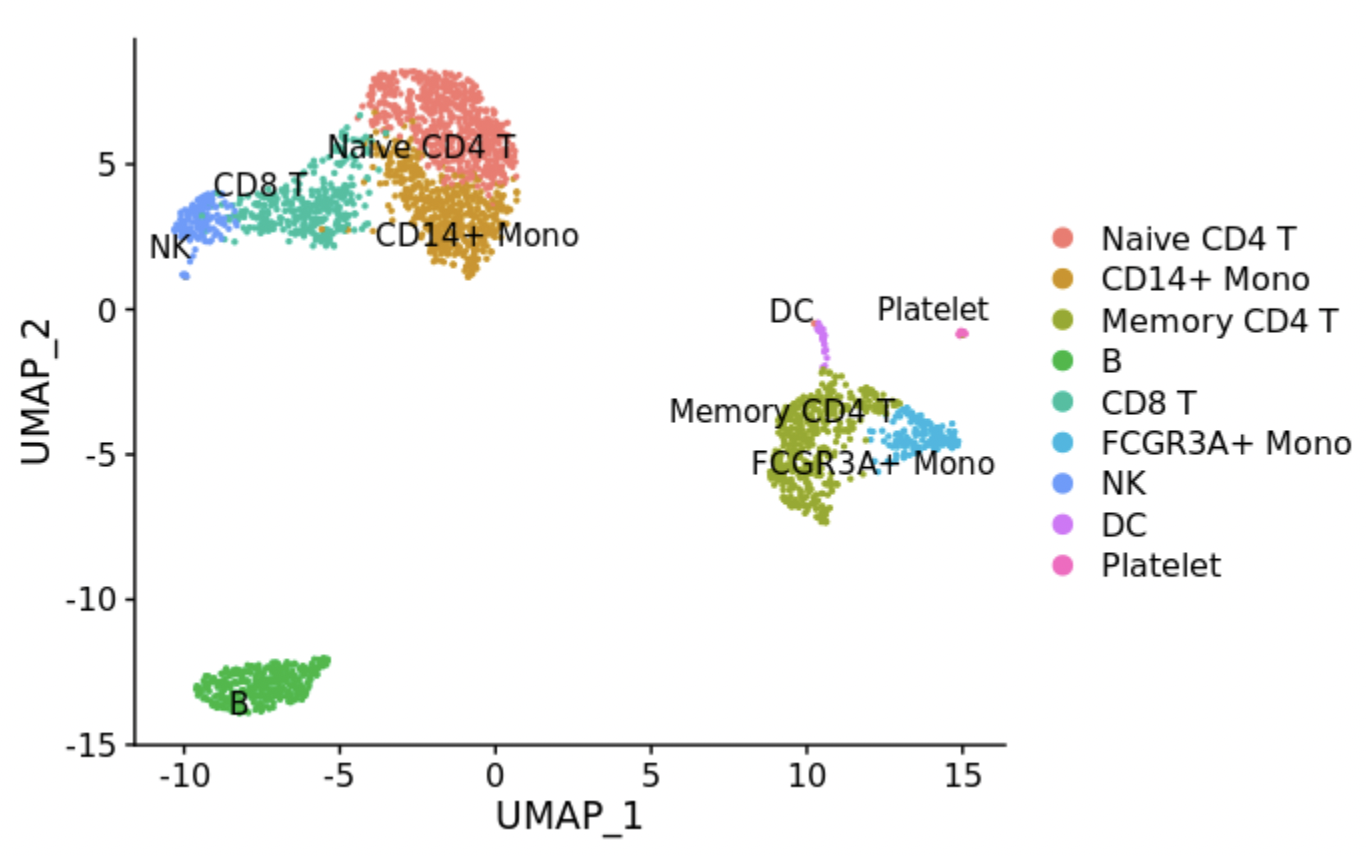}
		\caption{}\label{fig:2a}		
	\end{subfigure}
	\hspace{0cm}
	\begin{subfigure}[t]{3.5in}
		\includegraphics[width=3.5in]{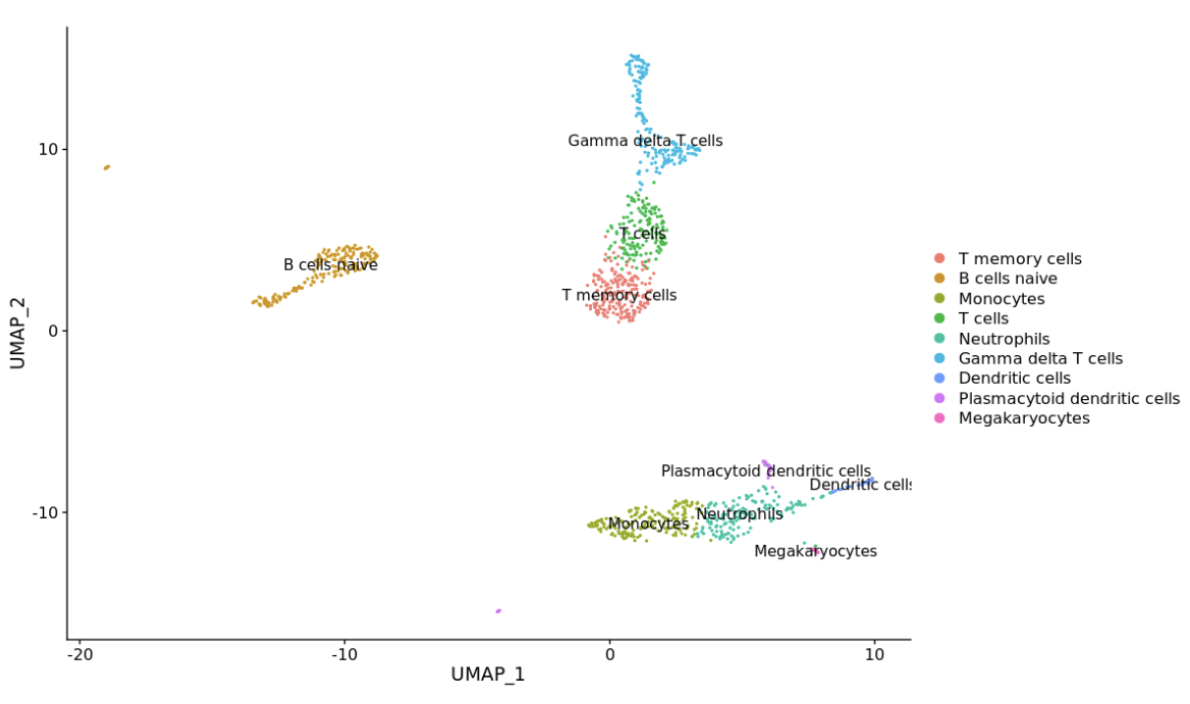}
		\caption{}\label{fig:2b}
	\end{subfigure}
	\hspace{0cm}
	\begin{subfigure}[t]{3in}
		\includegraphics[width=3in]{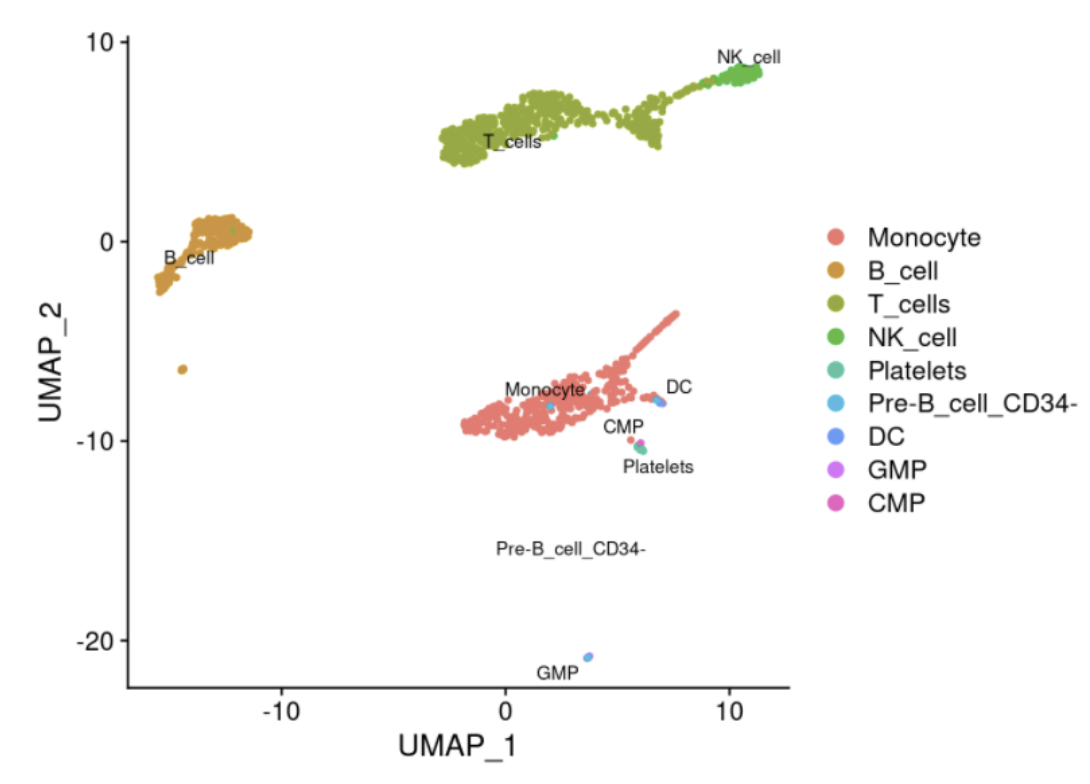}
		\caption{}\label{fig:2c}
	\end{subfigure}
	\hspace{0cm}
	\begin{subfigure}[t]{3in}
		\includegraphics[width=3.5in]{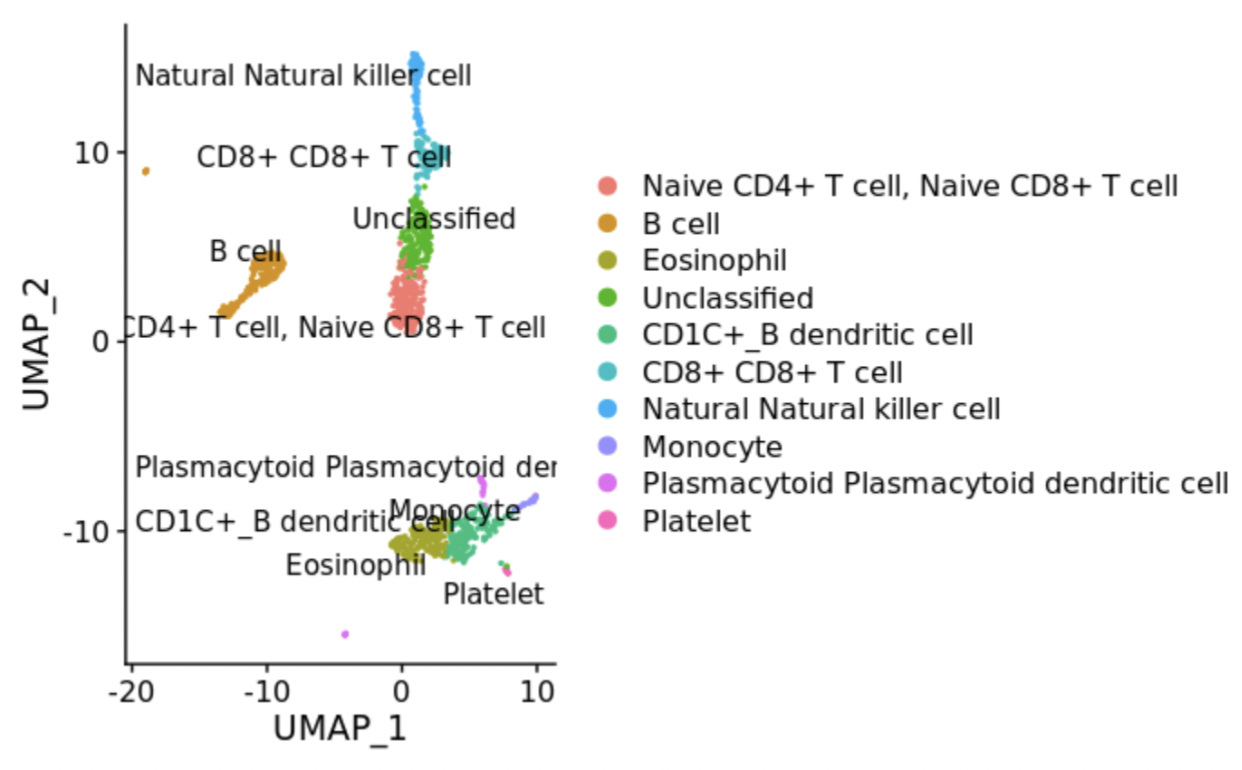}
		\caption{}\label{fig:2d}
	\end{subfigure}
	\hspace{0cm}
	\begin{subfigure}[t]{3in}
		\includegraphics[width=3.5in]{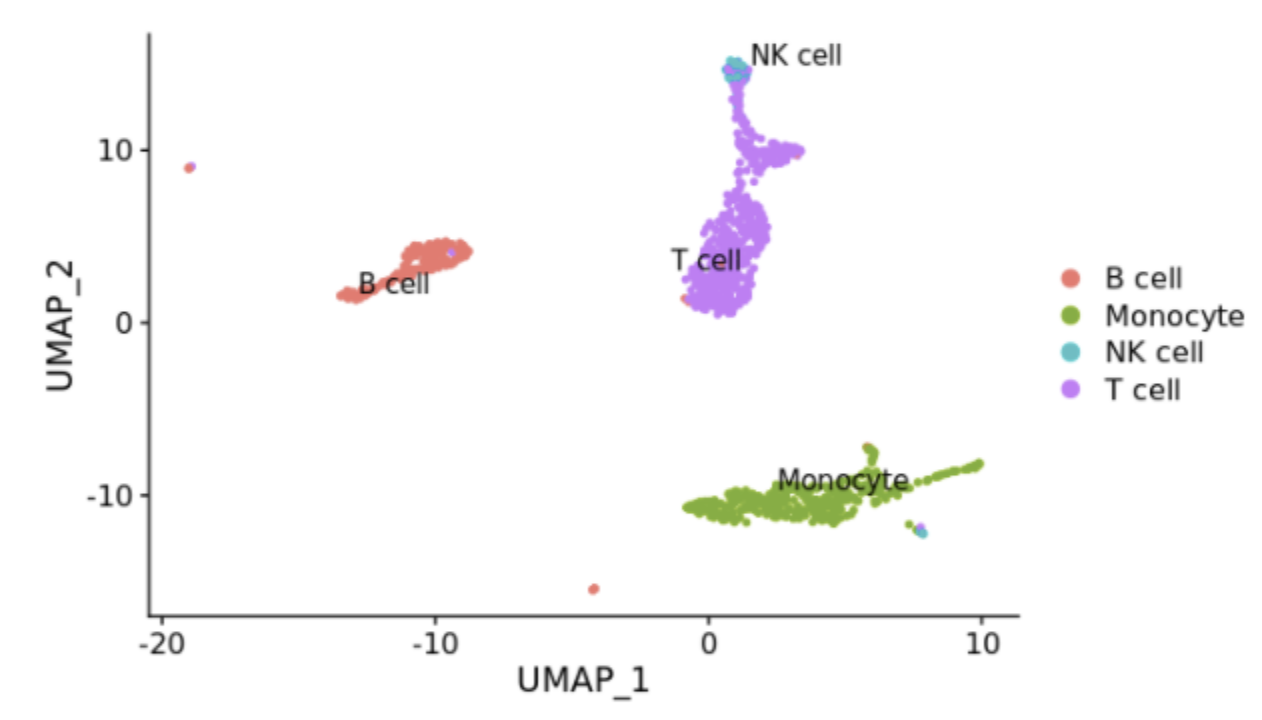}
		\caption{}\label{fig:2e}
	\end{subfigure}
	\caption{UMAP visualization of clustering by cell type, using (a) labels provided with the data set, (b) PanglaoDB sensitivity and specificity, (c) SingleR, (d) scCatch, and (e) ACTINN.}\label{fig:2}
\end{figure*}
\section{Results}

The current version of the pipeline successfully automates the process of downloading FASTQs from SRR run IDs, converting the FASTQ files into a gene expression matrix using Cell Ranger or Alevin, performing data preprocessing and clustering analysis using Seurat, and producing a HTML of the R Markdown file containing downstream visualizations and analyses. We used qualitative and quantitative techniques to validate the results from cell type identification methods.

\subsection{Visualization of Results Using the PBMC Dataset}
The PBMC dataset, which is expected to include an abundance of immune cell types, was used as the benchmark to evaluate the results generated by each method. The prediction labels were obtained from the baseline Seurat method under the published Seurat vignette, which serve as the ground truth values in comparing against the labels of the PanglaoDB, SingleR, scCATCH, and ACTINN methods for the purpose of this project. To evaluate each method qualitatively, we produced UMAP plots to visualize cell types (Fig. 2).

\subsection{Quantitative Evaluation of the Various Methods against the Original Seurat Method}
We examined the composition of the data set for each method by calculating the percentages of cells classified as each cell type. The five most common cell types for each method are as follows: 

\begin{itemize}
 \item Baseline Seurat: Naive CD4+ T cells (23.9\%), CD14+ Monocytes (18.8\%), Memory CD4+ T cells (18.1\%), B cells (13.0\%), CD8+ T cells (12.6\%);
 \item PanglaoDB: T memory cells (18.2\%), B cells naive (16.6\%), Monocytes (15.5\%), T cells (14.2\%), Neutrophils (13.6\%);
 \item SingleR: T cells (42.8\%), Monocytes (31.0\%), B cells (16.2\%), Natural Killer cells (4.78\%), platelets (1.0\%);
 \item scCATCH: Naive CD4+ T cells/Naive CD8+ T cells (18.2\%), B cells (16.56\%), Eosinophils (15.5\%), Unknown (14.2\% ), CD1C+\_B Dendritic cells (13.6\%); 
 \item ACTINN: T cells (45.3\%), Monocytes (33.6\%), B cells (17.7\%), Natural Killer (NK) cells (3.45\%).
\end{itemize}

All methods identified T cells (and subtypes of T cells) as the most common cell type in the PBMC dataset. All methods except scCATCH identified monocytes as the second most common cell types. However, the composition varied among all methods for less common cell types.

\begin{figure}[htbp]
\centerline{\includegraphics[scale=0.6]{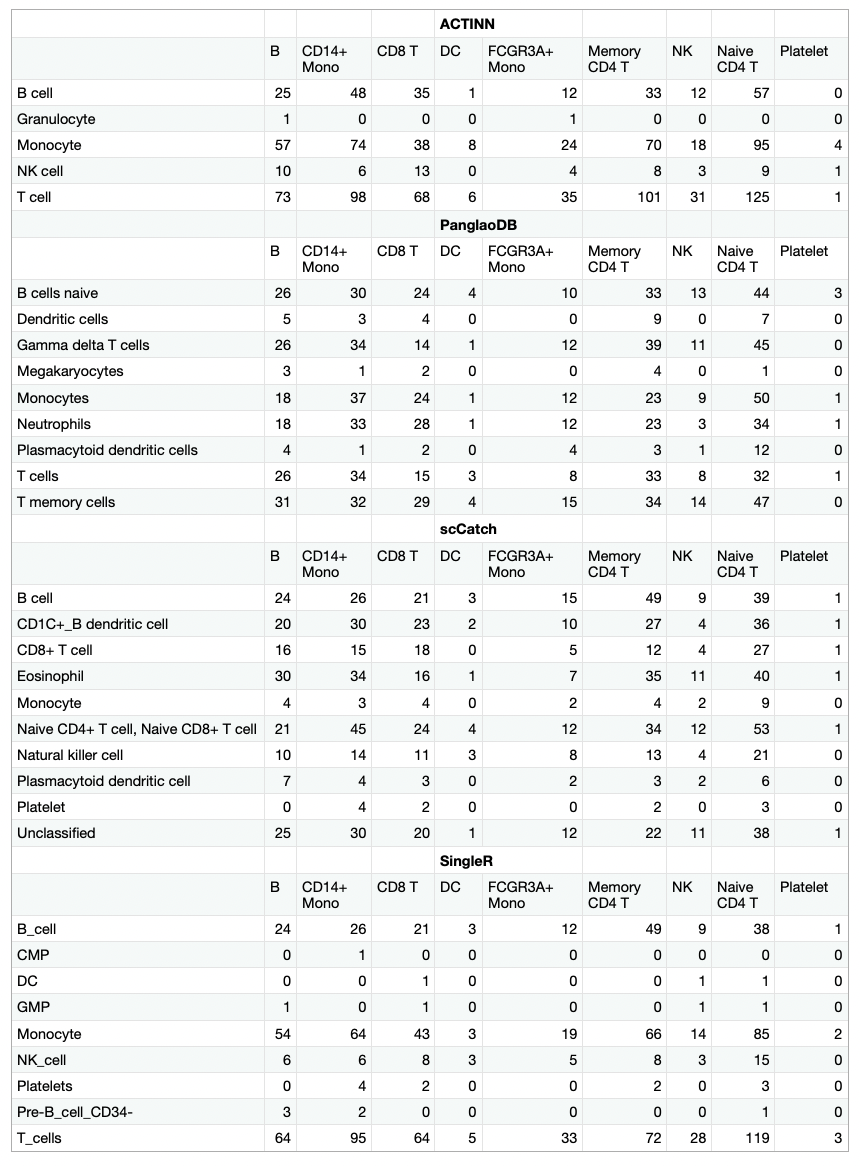}}
\caption{Confusion matrices for comparing the predictions of each cell type identification method (ACTINN, PanglaoDB, scCATCH, and SingleR) with the labels provided by Seurat.}
\label{fig}
\end{figure}

In addition, confusion matrices (Fig. 3) were used to compare the
prediction of each cell type identification method with the labels provided by the baseline Seurat method. In the matrices, the columns represent Seurat's labels and the rows represent the predictions by the various identification methods. Although the classification of individual cells differs considerably, this is in part due to distinct reference datasets used by each method, which differ in terms of the canonical makers and cell types included.

\section{Discussion}

According to Xie et al., Seurat, ACTINN, and SingleR all fall into the category of “eager learning” cell type identification methods \cite{b10}. This type of approach discerns highly variable genes and shared attributes of cells in the training data in order to group those cells by type then assign test cells to the most closely associated labeled group \cite{b10}. Though these and other methods placed in this category report high mean accuracy, their "unlabeling rates" (i.e., failure to identify a cell’s type) are also higher than other approaches \cite{b10}.

On the other hand, scCATCH is a “marker learning” cell type identification method \cite{b10}. Although this differs from Seurat, ACTINN, and SingleR, the need for canonical markers for each cell type can also lead to a higher unlabeling rate due to the absence of genes that can clearly differentiate between cell types and subtypes. Therefore, comparing and ensembling these methods may reduce unlabeling rates and uncertainty in predictions. It would be greatly beneficial to use an ensemble of the methods and incorporate a hard-voting classification scheme to derive the best possible classification of cell type for each individual cell. 

Additionally, a major limitation of this pipeline was not having a valid cross-model evaluation metric for the prediction accuracy of ACTINN because the labels in the training data were retrieved from the results of the PanglaoDB method, which indicates that this was not the most accurate existing approach due to the lack of statistical and experimental validations to deriving the training labels. This problem can be addressed by obtaining ground truth labels for training ACTINN from flow-cytometry data rather than basing these labels on the clustering results. Furthermore, the training data used for ACTINN can be expanded to incorporate more specific cell subtypes, which could lead to further insights into cellular composition and potential intercellular communication in the tumor microenvironment. 

Despite the advances made in automatic cell type identification methods, their relative novelty and abundance can make it difficult for those without domain-specific expertise both in Immunology and Data Science to use the tools appropriately for quantitative evaluation and cell type identification of single cell RNA sequencing. Furthermore, even those with experience in both fields can benefit from a standardized workflow for analyzing data, in particular when collaborating with others. The pipeline we created and the accompanying documentation provides researchers with an accessible and streamlined approach to ingesting and processing data, along with performing quality control measures, initial exploratory data analysis, and finally cell type identification. These results can be incorporated into studies on changes in gene expression and communication between immune and stromal cells with malignant cells, insights that can inform the development of therapeutics that target harmful mutations and abnormal signaling pathways.

\section{Conclusion}

The current pipeline can successfully automate the workflow from extracting scRNA-seq data to generating informative visualizations and predicting cell types. This pipeline can serve as the prototype for future work and expansion to include more features and enhance user-friendliness, such as including a more streamlined and interpretative approach to evaluating the cell type identification methods and increasing the automation of the pipeline. The current stage of the pipeline is limited to users who have access to Rivanna, which restricts the accessibility of the pipeline to only the researchers at the University of Virginia. Thus, enhancing the accessibility of the pipeline to a broader range of users is an important consideration in terms of driving the research of cell-cell communication and therapeutic innovation in the study of cancer.

\section*{Acknowledgments}

We would like to thank the UVA School of Data Science and the UVA Department of Biomedical Engineering. In particular, we would like to acknowledge Dr. Anne Talkington, Dr. Sepideh Dolatshahi, and Dr. Heman Shakeri for their generous guidance and support on this project.

\end{document}